\documentclass[aps,twocolumn,superscriptaddress,floatfix]{revtex4-2}
\usepackage{graphicx}
\usepackage{color}
\usepackage[version=4]{mhchem}
\definecolor{darkblue}{rgb}{0,0.02,0.45}
\usepackage[colorlinks=true,citecolor=dark blue,linkcolor=black]{hyperref}

\usepackage{multirow}
\usepackage{xfrac}
\usepackage{standalone}
\usepackage{ulem}
\usepackage{tikz}

\usepackage{newtxtext,newtxmath}

\newcommand\inputpgf[2]{{
		\let\pgfimageWithoutPath\pgfimage
		\renewcommand{\pgfimage}[2][]{\pgfimageWithoutPath[##1]{#1/##2}}
		\let\includegraphicsWithoutPath\includegraphics
		\renewcommand{\includegraphics}[2][]{\includegraphicsWithoutPath[##1]{#1/##2}}
		\input{#1/#2}
}}

\everymath=\expandafter{\the\everymath\displaystyle}

\AtBeginDocument{\catcode`\#=12}

\begin{document}

This manuscript has been authored by UT-Battelle, LLC, under Contract No.
DE-AC0500OR22725 with the U.S. Department of Energy. The United States
Government retains and the publisher, by accepting the article for publication,
acknowledges that the United States Government retains a non-exclusive, paid-up,
irrevocable, world-wide license to publish or reproduce the published form of this
manuscript, or allow others to do so, for the United States Government purposes.
The Department of Energy will provide public access to these results of federally
sponsored research in accordance with the DOE Public Access Plan (http://energy.gov/
downloads/doe-public-access-plan).
\clearpage

\title{Contrasting structural reversibility and magnetic correlations in isostructural honeycomb magnets CrCl$_3$ and $\alpha$-RuCl$_3$}

\author{Zachary Morgan}
\email{morganzj@ornl.gov}
\affiliation{Neutron Scattering Division, Oak Ridge National Laboratory,  Oak Ridge, Tennessee 37831, USA}

\author{Iris Ye}
\affiliation{Next Generation Pathway to Computing Program Participant}

\author{Jiasen Guo}
\affiliation{Neutron Scattering Division, Oak Ridge National Laboratory,  Oak Ridge, Tennessee 37831, USA}

\author{Michael A McGuire}
\affiliation{Materials Science and Technology Division, Oak Ridge National Laboratory, 
Oak Ridge, Tennessee 37831, USA}

\author{Jiaqiang Yan}
\affiliation{Materials Science and Technology Division, Oak Ridge National Laboratory, 
Oak Ridge, Tennessee 37831, USA}
\date{\today}

\begin{abstract}
We report a comparative neutron single crystal diffraction study of the structural and magnetic properties of layered halides CrCl$_3$ and $\alpha$-RuCl$_3$, which host a honeycomb arrangement of transition metal ions with distinct electronic configurations and undergo a first-order structural transition between high-temperature \textit{C}2/\textit{m} and low-temperature \textit{R}$\bar{3}$. Both compounds show a step-like change in the $c$-lattice, consistent with an expected stacking rearrangement. In contrast, the in-plane lattice response is quite different: $\alpha$-RuCl$_3$ exhibits an abrupt hysteretic change across the transition accompanied by progressive crystalline degradation upon thermal cycling, whereas CrCl$_3$ shows a smooth in-plane lattice evolution and remains structurally robust. Magnetically, CrCl$_3$ orders into an A-type antiferromagnetic structure at T$_N$=14\,K and exhibits pronounced diffuse magnetic scattering extending up to about 40\,K. $\alpha$-RuCl$_3$ shows no observable magnetic diffuse scattering above its zig-zag antiferromagnetic ordering temperature T$_N$=7.6\,K. These results suggest that the contrasting structural and magnetic behaviors arise from an interplay between interlayer sliding energetics and the fundamentally different electronic configurations of the two compounds.

\end{abstract}

\maketitle

\section{Introduction}

Layered, cleavable transition-metal trihalides ($MX_3$, $M=\mathrm{metal}$, $X=\mathrm{halide}$) have emerged as an important platform for exploring low-dimensional magnetism and correlated quantum phenomena \cite{burch_magnetism_2018,gong_two-dimensional_2019,trebst_kitaev_2022}. This class of materials hosts a wide range of emergent behaviors, including proximate Kitaev quantum spin liquid physics in $\alpha$-RuCl$_3$ \cite{plumb2014alpha, banerjee_proximate_2016}, layer-dependent ferromagnetism in CrI$_3$ \cite{huang_layer-dependent_2017}, topological magnon excitations in Cr$X_3$ compounds \cite{chen_topological_2018,cai_topological_2021}, and the spiral spin liquid in FeCl$_3$ \cite{gao_spiral_2022}. These diverse phenomena arise from the interplay between crystal symmetry, exchange interactions, and spin–orbit coupling \cite{mcguire_crystal_2017}.

A defining feature of these materials is the weak interlayer van der Waals coupling that allows multiple stacking sequences separated by small energetic barriers. As a result, several compounds exhibit temperature-driven structural transitions between nearly degenerate stacking arrangements even below room temperature \cite{mcguire_coupling_2015, banerjee_proximate_2016, kratochvilova2022crystal, park_emergence_2024}. These stacking configurations modify the symmetry and magnetic exchange pathways and influence the magnetic ground state and collective excitations. Understanding the relationship between stacking structure and magnetism as well as other physical properties remains a central challenge for layered halide magnets.

Motivated by recent work in $\alpha$-RuCl$_3$ that examined the influence of stacking disorder on its structural transition, magnetic properties, and thermal transport \cite{zhang_stacking_2024}, we explore how stacking geometry and structural stability relate to magnetic behavior. As a prototypical system with strong spin–orbit coupling and anisotropic exchange interactions, $\alpha$-RuCl$_3$ exhibits a pronounced sensitivity of its physical properties to structural defects. However, it remains unclear whether this is simply related to interlayer interactions connected to the stacking degree of freedom, or to a coupling between the stacking sequence and the local electronic degrees of freedom.

\begin{figure}[!htbp]
   \resizebox{\linewidth}{!}{\input{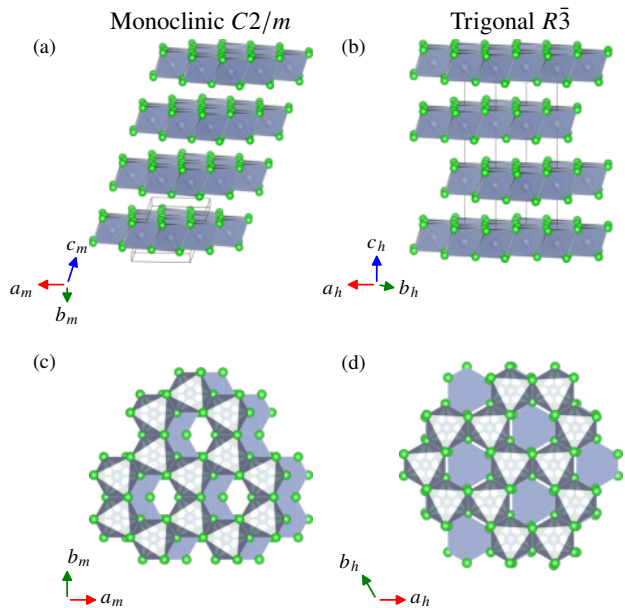}}
   \caption{
   Stacking sequence of honeycomb layers with transition metal $M$Cl$_6$ octahedra ($M=\mathrm{Cr,Ru}$) in (a) high-temperature monoclinic $C2/m$ and (b) low-temperature trigonal $R\bar{3}$ structures. The honeycomb layers are composed of edge-sharing octahedra with distinct alignments: (c) the monoclinic phase consists of layers shifted in-plane along the $a$-axis, while (d) the trigonal phase exhibits a staggered stacking sequence in which metal ions reside above the centers of the honeycomb plaquettes in adjacent layers. The hexagonal axes with rhombohedral centering are used to describe the trigonal phase with a relation $\boldsymbol{a}_m=\boldsymbol{a}_h$, $\boldsymbol{b}_m=\boldsymbol{a}_h+2\boldsymbol{b}_h$, and $\boldsymbol{c}_m=(-\boldsymbol{a}_h+\boldsymbol{c}_h)/3$. In this convention, directions of the $a$-axes from both unit cells are aligned, however, $\boldsymbol{a}_m$ could equivalently be referenced along $\boldsymbol{b}_c$ or $-(\boldsymbol{a}_c+\boldsymbol{b}_c)$ via three-fold symmetry.
   }\label{struct}
\end{figure}

To disentangle these effects, it is useful to consider a structurally analogous compound with a fundamentally different electronic interactions. CrCl$_3$ provides such a reference: it shares the same layered honeycomb framework and undergoes a similar temperature-driven transition between the high-temperature monoclinic ($C2/m$) and low-temperature trigonal ($R\bar{3}$) phases. Figure\,~\ref{struct}(a) shows a schematic of the monoclinic stacking sequence of the honeycomb layer, characterized by shift along the $c$-direction of the unit cell which is not normal to the layer. As By contrast, Fig.~\ref{struct}(b) shows the trigonal stacking sequence with the shift $[\sfrac{2}{3},\sfrac{1}{3},\sfrac{1}{3}]$ in the hexagonal setting (equivalent to the rhombohedral setting $\boldsymbol{a}_R$-direction). Two adjacent layers viewed normal to the honeycomb network are shown in Fig.~\ref{struct}(c) and (d), highlighting the distinct stacking configurations of the monoclinic and trigonal phases, respectively. It can be understood that the monoclinic phase consists of layers appear shifted in-plane along the $a$-axis, whereas in the trigonal phase the layers are shifted such that the metal ions are positioned above the honeycomb voids of neighboring layers.

This first-order structural transition occurs at approximately 140~K (170~K) in $\alpha$-RuCl$_3$ and 240~K (260~K) in CrCl$_3$ upon cooling (warming). The lower transition in $\alpha$-RuCl$_3$ indicates either a smaller energy difference between the competing stacking sequences or a larger entropy of the high-temperature phase. Despite these structural similarities, the two compounds host fundamentally different electronic and magnetic interactions. In CrCl$_3$, the 3d$^3$ electronic configuration gives rise to predominantly isotropic Heisenberg exchange and conventional Heisenberg-like magnetism consisting of ferromagnetic layers that couple antiferromagnetically \cite{cable_neutron_1961}. In contrast, $\alpha$-RuCl$3$ is a proximate Kitaev quantum spin liquid in which spin-orbit-entangled $J\mathrm{eff}=1/2$ states produce strongly anisotropic bond-dependent interactions and a zigzag antiferromagnetic ground state \cite{sears_magnetic_2015, johnson_monoclinic_2015, banerjee_proximate_2016, banerjee_neutron_2017, lang2016unconventional}. A direct comparison between CrCl$_3$ and $\alpha$-RuCl$_3$ therefore provides a route to disentangle the roles of generic lattice stacking from that of underlying electronic configurations.

In this work, we present such a comparative study of the structural and magnetic properties of CrCl$_3$ and $\alpha$-RuCl$_3$ using single-crystal neutron diffraction. The structures of both CrCl$_3$ and $\alpha$-RuCl$_3$ have been extensively studied in the literature, here we focus on possible diffuse magnetic scattering and the effects of thermal cycling on the structural reversibility. Our results show that repeated thermal cycling across the structure transition degrades the crystalline quality of $\alpha$-RuCl$_3$, whereas CrCl$_3$ remains structurally robust. This difference is likely associated with the step-like, hysteretic in-plane lattice response to the structure transition observed in  $\alpha$-RuCl$_3$ that is absent in CrCl$_3$. Magnetically, CrCl$_3$ orders at $T_N = 14$~K into ferromagnetic layers aligned within the basal plane that couple antiferromagnetically between layers and exhibits pronounced diffuse magnetic scattering extending up to approximately 40~K, indicating persistent short-range spin correlations. $\alpha$-RuCl$_3$ shows no observable magnetic diffuse scattering above its ordering temperature within the sensitivity of our measurements. Our results suggest that in addition to interlayer sliding energetics, the electronic configuration also plays an important role for the contrasting behavior. The observed crystalline degradation in $\alpha$-RuCl$_3$ after thermal cycling further complicates the experimental realization of the half-integer quantized thermal Hall effect and should be considered in future studies. 

\section{Experimental Details}

Single crystals used in this study were grown using self-selecting vapor transport method \cite{yan_self-selecting_2023}. The masses of the crystals used for diffraction measurements ranged from 60 to 200\,mg. The magnetic properties of crystals grown in the same growth condition have been well characterized in previous work \cite{yan_self-selecting_2023, zhang_stacking_2024}. Neutron diffraction experiments were conducted at the Spallation Neutron Source using the time-of-flight diffuse scattering spectrometer CORELLI~\cite{ye_implementation_2018}. Each crystal was mounted on an aluminum pin using super glue, with particular care taken to avoid applying mechanical stress to the sample. The crystal was then loaded into a closed-cycle refrigerator with a base temperature of 5.5~K and aligned on a single-axis goniometer.

To investigate the nuclear and magnetic structures, mesh scans were performed in 2$^\circ$ rotational increments with an acquisition time of 1.5~min/angle, covering a full 360$^\circ$ rotation. Measurements were carried out at four temperatures: first, just above and below the structure transition; another set near and abovFe the magnetic ordering temperature to investigate possible magnetic diffuse scattering, and at base temperature in the magnetically ordered phase. 

To further probe the structure transition, the sample was held in a fixed orientation while tracking the temperature-dependent intensity of a selected nuclear Bragg peak that is allowed in the monoclinic phase but disappears in the trigonal phase. Temperature ramps were performed to investigate the hysteresis at a ramp rate of 2~K/min.

At base temperature, a magnetic Bragg peak with characteristic wavevector of the magnetic structure was located. Orienting the crystal to optimize the signal-to-noise ratio, the thermal evolution of the Bragg peak intensity was tracked upon warming at a rate of 0.5~K/min. After observing diffuse scattering in CrCl$_3$, the temperature dependence along the $[0,0,l]$ was tracked by rocking the crystal $\pm$14$^\circ$ in 2$^\circ$ increments, 2~min/angle held at discrete temperatures up to 39~K.

\section{Results}

\begin{figure}[!htbp]
   \resizebox{\linewidth}{!}{\input{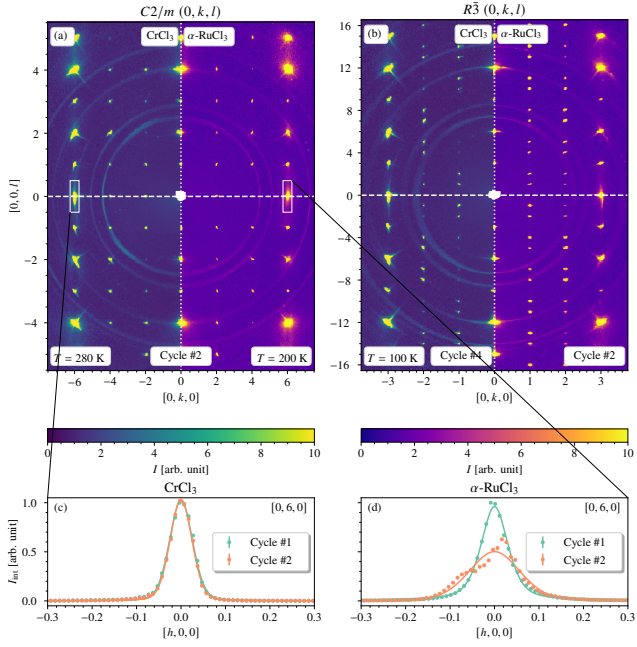}}
   \caption{Crystalline degradation upon thermal cycling occurs in $\alpha$-RuCl$_3$ but not in CrCl$_3$ determined from neutron diffraction. The reciprocal space maps of the (a) monoclinic $(0,k,l)$ and (b) trigonal $(h,0,l)$ scattering planes after thermal cyclings without any clear signature of structural diffuse scattering due to stacking order. Above the dashed line indicate the pristine state with the cycled state shown below. Left and right of the dashed line correspond to CrCl$_3$ and $\alpha$-RuCl$_3$, respectively. Comparison of the peak width of the $[0,6,0]$ monoclinic ($[0,3,0]$ hexagonal) reflection in-plane after cycling for (c) CrCl$_3$ and (d) $\alpha$-RuCl$_3$. Solid line is a fit to a Lorentzian-like profile.
   }\label{transition}
\end{figure}

\begin{figure}[!htbp]
   \resizebox{\linewidth}{!}{\input{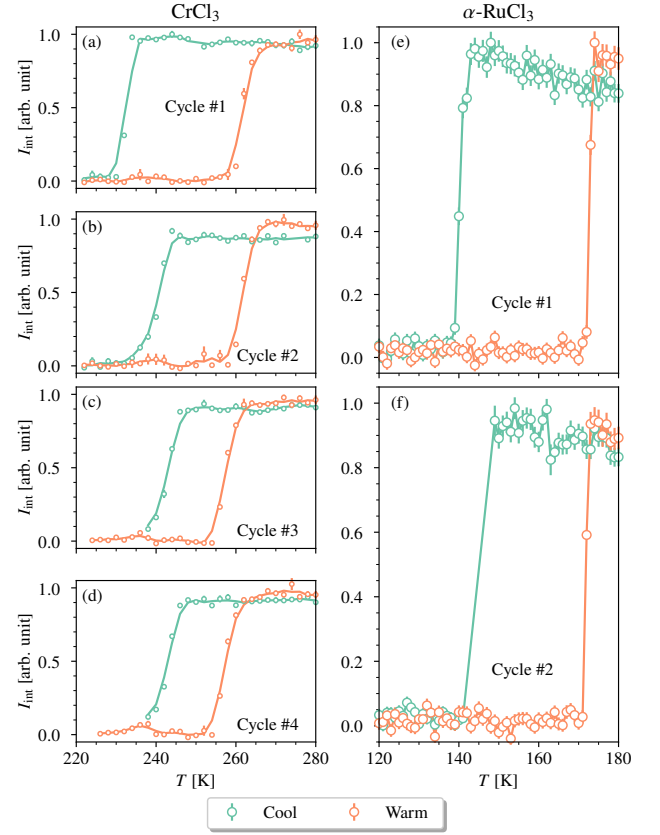}}
   \caption{The hysteresis loops of CrCl$_3$ (a-d) first through fourth thermal cycles showing the integrated intensity $I_\mathrm{int}$ of the high-temperature $[2,0,1]$ peak forbidden in the low-temperature structure. For $\alpha$-RuCl$_3$, the (e,f) first and second cycles show the $[1,1,6]$ peak. Solid lines are guides to the eye.
   }\label{hysteresis}
\end{figure}

\begin{figure}[!htbp]
   \resizebox{\linewidth}{!}{\input{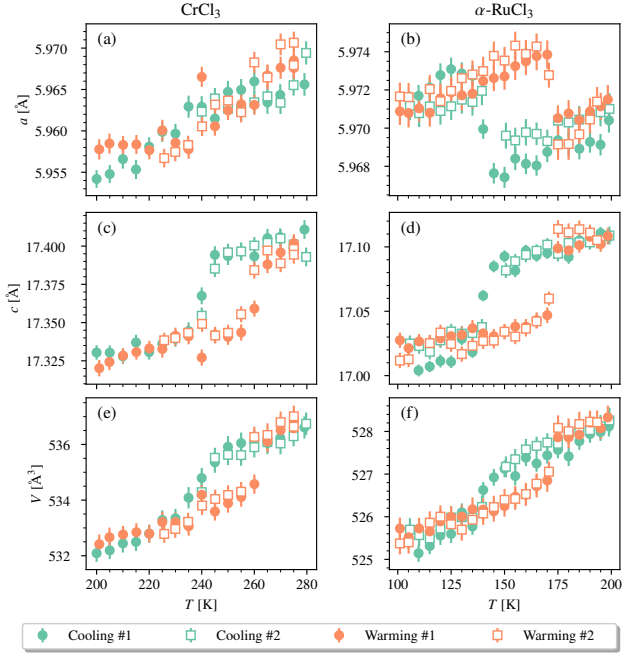}}
   \caption{Comparison of the lattice constant (a,b) $a_h$, (c,d) $c_h$, and (e,f) unit cell volume change across the structure transition in CrCl$_3$ and in $\alpha$-RuCl$_3$, calculated using 95 and 126 reflections, respectively. Two thermal cycles are overlaid showing consistent behavior.
   }\label{lattice_constants}
\end{figure}

\subsection{Structural response to thermal cycling}

The structure and its high- to low-temperature transition in both compounds have been extensively studied using diffraction, spectroscopy, bulk measurements, and first-principles calculations \cite{johnson_monoclinic_2015,kim_crystal_2016,cao_low-temperature_2016,mcguire_magnetic_2017,kratochvilova_crystal_2022,mu_role_2022,sears_stacking_2023,park_emergence_2024,zhang_2d_2024,morgan_structure_2024,schneeloch_role_2024,hong_first-principles_2026, kim2024structural, cao_low-temperature_2016}. Our measurements confirm previously reported structural transition from $C2/m$ to $R\bar{3}$ upon cooling \cite{mcguire_magnetic_2017,mu_role_2022,sears_stacking_2023,park_emergence_2024,zhang_2d_2024,morgan_structure_2024,schneeloch_role_2024,kim2024structural}. Here, we focus on the evolution of the structure under thermal cycling.

Figures~\ref{transition}(a) and (b) show the representative reciprocal-space maps of the monoclinic and trigonal phases, respectively, for both compounds displaying the distinct stacking signatures of the high- and low-temperature structures. The temperature induced transition between these phases, shown in Fig.~\ref{hysteresis}, is of first-order with thermal hysteresis. The transition of the pristine (first cooling) CrCl$_3$ shown in Fig.~\ref{hysteresis}(a) occurs at $\sim260$~K upon warming and $\sim240$~K upon cooling. Figures~\ref{hysteresis}(b)–(c) track the hysteresis across successive cycles in CrCl$_3$. The structure transition in $\alpha$-RuCl$_3$, shown for two cycles in Figs.~\ref{hysteresis}(e) and (f), occurs at $\sim170$~K upon warming and $\sim140$~K upon cooling. This indicates that the $\alpha$-RuCl$_3$ crystals used in this study are of high quality following the materials selection criteria previously developed \cite{zhang_stacking_2024}. 

Given the first-order nature of this transition, we examine the effect of repeated thermal cycling. CrCl$_3$ appears to show no observable degradation upon thermal cycling. Reciprocal-space features and peak widths remain unchanged, while the hysteresis loop narrows slightly with cycling. The instrumental (Laue) view indicates no measurable change in mosaic or introduction of extra domains, and the extracted structure factors remain reproducible across cycles in both phases \cite{supp_mat}.

$\alpha$-RuCl$_3$, on the other hand, exhibits degradation of crystalline quality. While the reciprocal-space maps indicate minimal change in mosaic spread along the out-of-plane $[0,0,l]$ direction as shown in Figs~\ref{transition}(a)-(b), line cuts through the $[0,6,0]$ Bragg peak along the in-plane direction $[h,0,0]$ [Figs.~\ref{transition}(c) and (d)] reveal significant peak broadening after cycling compared to CrCl$_3$. The corresponding correlation length, estimated from the full width at half maximum (FWHM), increases slightly in CrCl$_3$ from 94(1) to 99(1)~\AA, but decreases significantly in $\alpha$-RuCl$_3$ from 78(1) to 42(1)~\AA, indicating a reduction of in-plane coherence.

The out-of-plane correlation length of CrCl$_3$ remains essentially unchanged upon thermal cycling, varying only weakly from 250(3)~\AA in the pristine state to 236(2)~\AA\ after the first cycle and 241(2)~\AA\ after the fourth cycle $\alpha$-RuCl$_3$ exhibits a modest reduction in out-of-plane coherence, with the correlation length decreasing from 342(2)~\AA in the pristine state to 312(1)~\AA after the first thermal cycle, consistent with our previous study \cite{zhang_stacking_2024}.

Additionally, the Laue instrument view shows an increase in mosaic spread for $\alpha$-RuCl$_3$, and the integrated intensities exhibit increased reflection-to-reflection variation after cycling, despite comparable crystallographic $R$-factors \cite{supp_mat}. In the trigonal phase, there is a repopulation of obverse and reverse twin fractions which contributes to the integrated intensity variation where it is essentially unchanged in CrCl$_3$.

To further compare these compounds, the lattice evolution across the transition is tracked by obtaining the centroids of reflections common to both $C2/m$ and $R\bar{3}$ phases as a function of temperature. The lattice parameters $a_h$ and $c_h$ are refined in each temperature bin using a consistent set of reflections down to $d_{\min}=0.4$~\AA. The reflections transformed into the hexagonal setting such that indexing follows $h_h = h_m$, $k_h = (k_m-h_m)/2$, and $l_h = 3l_m+h_m$ where $h=\mathrm{hexagonal}$ and $m=\mathrm{monoclinic}$ unit cell \cite{park_emergence_2024}. In this convention, $a_h$ describes the honeycomb edge length and $c_h$ the interlayer spacing. The resulting lattice evolution with temperature across the structure transition is summarized in Fig.~\ref{lattice_constants}.

The structural transition in these two compounds is associated with a rearrangement of honeycomb layers into distinct stacking sequences. In the low-temperature structure, the layers adopt a different stacking registry that modifies the relative arrangement of the halide networks across the van der Waals gap. In particular, the halide triangles between adjacent layers become more compactly nested compared to the high-temperature monoclinic phase, leading to a reduction of the $c_h$ lattice parameter upon cooling through the structural transition. Because of the weak interlayer van der Waals interactions, this layer rearrangement is not expected to induce significant intralayer lattice response. Therefore, upon cooling, the intralayer honeycomb lattice parameters are expected to evolve smoothly, while the interlayer spacing undergoes a more step-like reduction. 

This is indeed the case for CrCl$_3$. As shown in Figure~\ref{lattice_constants},  $a_h$ of CrCl$_3$ evolves smoothly through the transition, and the interlayer spacing $c_h$ displays a clear step-like contraction upon cooling and expansion upon heating in both systems. However, $\alpha$-RuCl$_3$ exhibits an abrupt, hysteretic change in both its in-plane and out-of-plane lattice parameters. The change of the in-plane lattice is about 0.05\%, smaller than 0.4\% change in $c_h$-lattice. As shown in Fig.~S5 in the supplemental material, this in-plane lattice response is evident looking at corresponding in-plane Bragg peaks that are insensitive to the layer spacing \cite{supp_mat}. A similar in-plane lattice change is also observed in $\alpha$-RuCl$_3$ crystals with T$_N$=6.5\,K and 10\%-Ir substituted $\alpha$-RuCl$_3$ \cite{supp_mat}. As discussed later, the unexpected in-plane lattice response might contribute to the crystal degradation upon thermal cycling and play an important role in the sensitivity of its bond dependent interactions to local bonding geometry. The unit-cell volume $V_h=(\sqrt{3}/2)a_h^2c_h$ [Figs.~\ref{lattice_constants}(e) and (f)] follows the same trend and is dominated by the change in $c_h$.

Given the observed in-plane lattice parameter change, it is natural to ask whether there are also corresponding changes in the local structure or the octahedral distortions. To probe the local structure, refinements were performed above and below the transition to extract atomic positions and displacement parameters \cite{supp_mat}. The distortion of the coordination environment was quantified using metrics derived from the spatial distribution of metal--ligand bond vectors surrounding each metal atom. These include the average bond length $d_{\mathrm{ave}}$, its standard deviation $d_{\mathrm{std}}$, and the normalized bond-length distortion $\Delta$, defined as the mean square of the relative deviation from the average bond length. Angular distortions of the ligand--metal--ligand geometry were further decomposed into 90$^\circ$- and 180$^\circ$-like variances. All of these parameters are summarized in the supplemental material \cite{supp_mat}.

In CrCl$_3$, the monoclinic phase exhibits a small but finite splitting of the Cr--Cl bond lengths, which is largely suppressed upon transition to the trigonal phase, while the angular distortions vary only modestly between $C2/m$ and $R\bar{3}$. $\alpha$-RuCl$_3$ retains nearly uniform Ru--Cl bond lengths in both phases, with only minor changes in bond-angle metrics. The evolution of the average bond lengths is consistent with the measured changes in the lattice parameters, although the refinements were performed at temperatures spanning the entire hysteresis window of the structural transition. Importantly, the refined local structure parameters remain unchanged within experimental uncertainty before and after thermal cycling. This indicates that the crystalline degradation observed in $\alpha$-RuCl$_3$ arises from the accumulation of lattice defects affecting long-range structural coherence rather than from irreversible modifications of the local coordination environment. This behavior is consistent with the reproducibility of the refined structure factors under thermal cycling, where CrCl$_3$ remains essentially unchanged \cite{supp_mat}.

\begin{figure*}[!htbp]
   \resizebox{\linewidth}{!}{\input{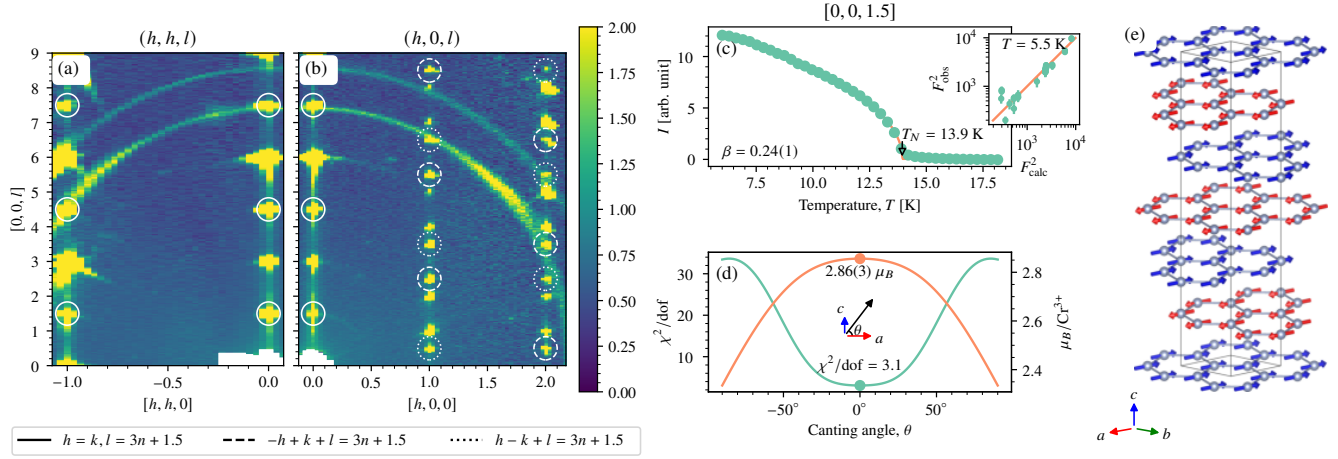}}
   \caption{
Base-temperature ($T=5.5$~K) neutron diffraction of CrCl$_3$. (a) the $(h,h,l)$ and (b) the $(h,0,l)$ scattering planes, showing the appearance of magnetic Bragg peaks at offset $[0,0,\sfrac{3}{2}]$ (circled). In rhombohedral space groups such as $R\bar{3}$, obverse and reverse twin domains are present, with nuclear Bragg peaks following $-h+k+l=3n$ and $h-k+l=3n$ reflection conditions, respectively. Contributions from each domain are circled, including overlapping reflections at $h=k$, such as $[\bar{1},\bar{1},l]$ and $[0,0,l]$.  
(c) The N\'eel temperature $T_N \approx 14$~K, determined from the integrated intensity of the $[0,0,\sfrac{3}{2}]$ peak, follows a critical exponent of $\beta \approx 0.24$ where the temperature-dependent intensity is $I(T)\propto(1-T/T_N)^{2\beta}$.
(d) Magnetic structure refinement at $T=5.5$~K of the canting angle away from the basal plane shows a minimum in the goodness-of-fit at $\theta=0$, though the curve is shallow with finite nonzero angles also possible. The inset compares observed and calculated structure factors for $\theta=0$, yielding an ordered moment of 2.86(3)~$\mu_B$. The magnetic domain populations refine to equal ratios once the obverse/reverse structural domain contributions (3:2) are accounted for, consistent with the nuclear structure refinement.  
(e) Low-temperature magnetic structure of \ce{CrCl3}, consisting of ferromagnetic ordering within single layers stacked in an alternating antiferromagnetic configuration between layers.
   }\label{mag}
\end{figure*}

\begin{figure*}[!htbp]
   \resizebox{\linewidth}{!}{\input{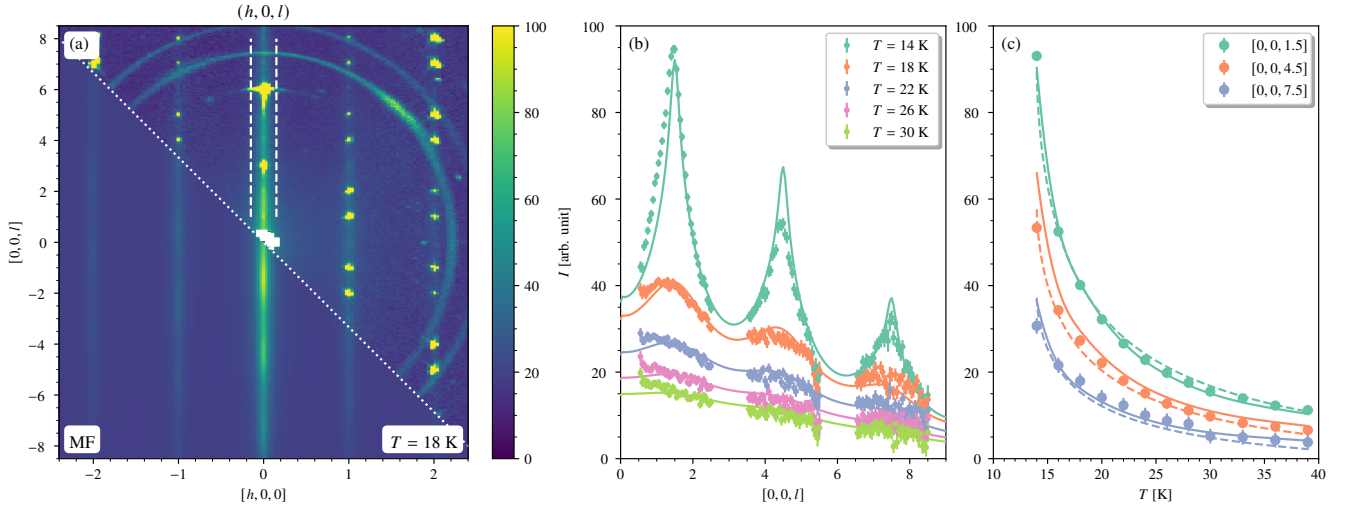}}
   \caption{
Magnetic diffuse scattering in CrCl$_3$. (a) Reciprocal space map of the CrCl$_3$ $(h,0,l)$ scattering plane at low-temperature showing elongated diffuse scattering features along $[0,0,l]$ at $T=18$ K above the antiferromagnetic ordering temperature. The upper-right of collected data is compared to the lower-left that shows a mean-field calculation of the spin Hamiltonian with weak antiferromagnetic interlayer coupling showing weaker features along $[\pm1,0,l]$ and $[\pm2,0,l]$ attenuated by the Cr$^{3+}$ form factor \cite{supp_mat}. 
(b) The $[0,0,l]$ profile at selected temperatures compared with the calculation showing good agreement.
(c) The magnetic diffuse scattering at $[0,0,\sfrac{3}{2}]$, $[0,0,\sfrac{9}{2}]$ , and $[0,0,\sfrac{15}{2}]$ persists up to 40 K. Solid line is calculation from best-fit Hamiltonian calculation.
   }\label{diff}
\end{figure*}

\begin{figure}[!htbp]
   \resizebox{\linewidth}{!}{\input{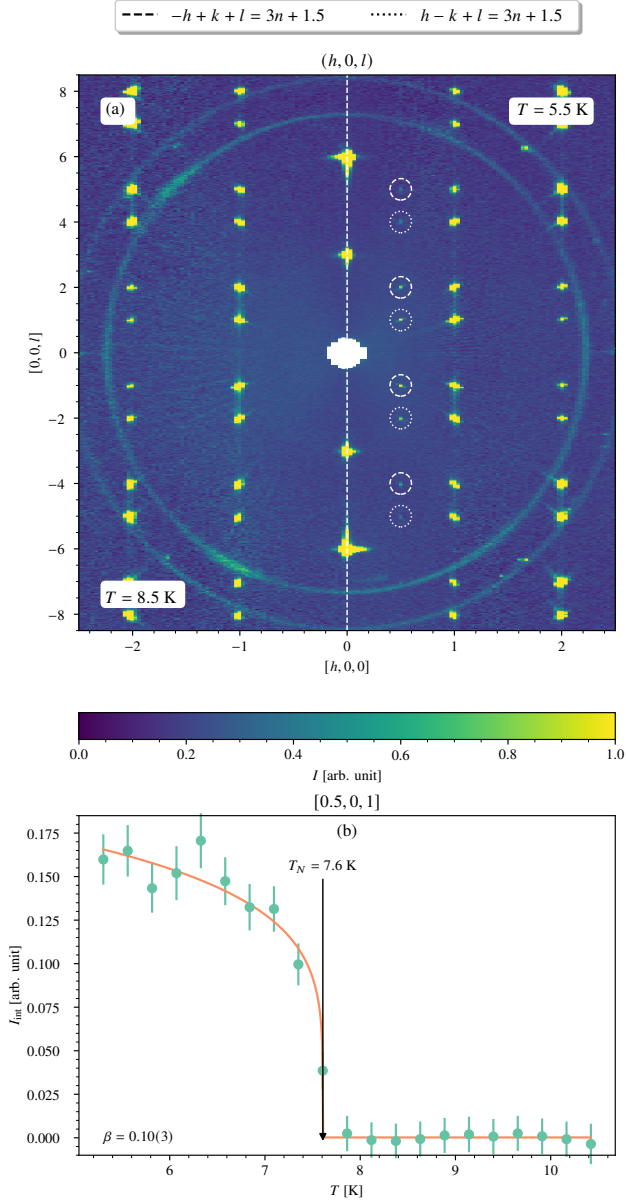}}
   \caption{
Absence of magnetic diffuse scattering in $\alpha$-RuCl$_3$. (a) Reciprocal-space map of $\alpha$-RuCl$_3$ in the $(h,0,l)$ scattering plane at low-temperature. 
The left panel shows the map at $T=8.5$\,K, just above the magnetic transition temperature $T_N=7.6$\,K, where no diffuse scattering features are observed. 
The right panel shows the map at base temperature ($T=5.5$\,K), where magnetic reflections appear at $[0,\sfrac{1}{2},1]$, circled, corresponding to the obverse and reverse twin domains. 
(b) Temperature dependence of the integrated intensity $I_\mathrm{int}$ of the $[0.5,0,1]$ reflection, showing two-dimensional Ising-like critical behavior. 
   }\label{magnetic_Ru}
\end{figure}

\subsection{Magnetic correlations above $T_N$}

Upon cooling CrCl$_3$ to the base temperature (5.5~K), additional Bragg reflections of magnetic origin appear at positions corresponding to the propagation vector $[0,0,\sfrac{3}{2}]$, situated between the nuclear reflections. These magnetic peaks originate from the obverse and reverse structural domains common in rhombohedral crystals \cite{herbst-irmer_refinement_2002}. This requires explicit consideration of their combined contributions when analyzing the observed scattering intensities. As illustrated in Fig.\ref{mag}(a), within the $(h,h,l)$ plane, the magnetic signals from both domains overlap, whereas in Fig.~\ref{mag}(b), corresponding to the $(h,0,l)$ plane, the respective contributions are clearly separated.

The temperature evolution of the magnetic $[0,0,\sfrac{3}{2}]$ Bragg peak intensity is shown in Fig.~\ref{mag}(c) with transition temperature around 14~K. At 5.5~K, the magnetic moment size is 2.86(3)$\mu_B$ close to early neutron work (2.7(1)-3.2(4)$\mu_B$ at 4.2~K) \cite{cable_neutron_1961}. The inset in panel Fig.~\ref{mag}(c) shows the refined magnetic structure factors to the observed ones. Magnetic symmetry analysis for propagation vector $[0,0,\sfrac{3}{2}]$ with Cr atom at the $6c$ Wyckoff position suggests that only one magnetic space group gives rise to the layered antiferromagnetic arrangement 2-$P_S-\bar{1}$ (\#2.7 BNS) \cite{perez-mato_symmetry-based_2015}. However, an out-of-plane component is symmetry allowed by the magnetic space group. Fig.~\ref{mag}(d) shows the goodness-of-fit as a function of an imposed out-of-plane canting angle showing a minimum at 0$^\circ$. For small angles, the fit is insensitive given by the shallow goodness-of-fit. The refined antiferromagnetic structure is illustrated in Fig.~\ref{mag}(e) with alternating ferromagnetic layers shown in the expanded magnetic cell along the $c$-axis of the hexagonal unit cell. 

Above the CrCl$_3$ magnetic transition, prominent diffuse scattering is observed with a distinct rod-like fingerprint features along $[0,0,l]$ as shown in Fig.~\ref{diff}(a). The diffuse scattering extends further out along the $c^\ast$-direction with broad humps centered beneath the propagation vector positions of $[0,0,\sfrac{3}{2}+3n]$ with $n$ being an integer. Weaker rods are also observed along $[1,0,l]$ and $[1,1,l]$. To model the diffuse scattering, mean-field calculations \cite{paddison_scattering_2020,gao_suppressed_2021} are performed using previously reported spin Hamiltonian exchanged interactions \cite{chen_massless_2021,do_gaps_2022,schneeloch_gapless_2022} as a starting point, increasing the interlayer coupling interaction $J_c$ to improve the fit along $[0,0,l]$ \cite{supp_mat}. Compared to previous estimates ($J_c=-0.001$ to -0.04~meV) based on spin wave analysis where the interlayer exchange parameter is less sensitive, the diffuse scattering analysis reveals a larger $J_c=-0.12(1)$~meV \cite{supp_mat}. The calculated intensity map is shown along side the observed data. Fig.~\ref{diff}(b) shows selected temperatures of the $[0,0,l]$ linecut with the calculation overlayed on top. For clarity, the strong nuclear Bragg peaks are removed from the observed data. The temperature dependence of the diffusive intensity is shown in Fig.~\ref{diff}(c) for selected regions corresponding to $n=0,1,2$. 

The magnetic structure of $\alpha$-RuCl$_3$ has been extensively studied using neutron diffraction \cite{ritter_zigzag_2016,morgan_structure_2024,park_emergence_2024}. Here, our investigation focuses on whether magnetic diffuse scattering is present. Figure~\ref{magnetic_Ru}(a) shows the reciprocal-space map measured at base temperature ($T=5.5$~K), where magnetic Bragg peaks with the characteristic propagation vector $[0,\sfrac{1}{2},1]$ are observed, confirming the presence of long-range magnetic order. The thermal evolution of the magnetic peak intensity, shown in Fig.~\ref{magnetic_Ru}(b), confirms an ordering temperature of $T_N \approx 7.6$~K and exhibits critical behavior consistent with a two-dimensional Ising universality class ($\beta=1/8$). This reflects the strong bond-directional (Kitaev) anisotropy, which constrains the spins and stabilizes zigzag order consisting of ferromagnetic chains along the honeycomb edges that are antiferromagnetically coupled between neighboring chains. About 1~K above $T_N$=7.6~K, inspection of slices through the reciprocal-space maps does not show magnetic diffuse scattering signature. This is in sharp contrast to the well resolved magnetic diffuse features in CrCl$_3$. 

Our refinement of the magnetic structure in $\alpha$-RuCl$_3$ is consistent with previously reported zigzag models based on unpolarized neutron diffraction \cite{ritter_zigzag_2016,cao_low-temperature_2016,park_emergence_2024}. However, it should be noted that a most recent neutron diffraction study was able to determine the full three dimensional orientation of the Ru moments by using both spherical and longitudinal neutron polarization analysis. Their results reveal that in the zigzag magnetic order the Ru$^{3+}$ magnetic moments tilt by 15.7$^{\circ}$ out of the basal plane and twist from the $a$-axis by -13.8$^{\circ}$ in the plane \cite{wang_tilted_2026}. 

\section{Discussion}

Both CrCl$_3$ and $\alpha$-RuCl$_3$ undergo a first-order structural transition between high-temperature ($C2/m$) and low-temperature ($R\bar{3}$) phases driven by weak interlayer interactions with nearly degenerate stacking configurations \cite{kim_crystal_2016,mcguire_crystal_2017}. Cooling through this transition, both compounds show a reduction of the interlayer spacing $c_h$, consistent with a rearrangement of the structural layers into an energetically favorable stacking sequence where halogen atoms reside above hollow sites of adjacent layers. These similarities suggest that the structural transition in both compounds is primarily associated with changes in the interlayer stacking sequence. 

Despite the above common features, these two compounds exhibit markedly different in-plane lattice responses and structural reversibility under thermal cycling. In CrCl$_3$, the in-plane lattice parameter evolves smoothly across the structure transition, indicating minimal perturbation of the honeycomb lattice by the stacking rearrangement\cite{havemann_comparison_2024}. Correspondingly, CrCl$_3$ remains structurally robust under repeated thermal cycling, with no observable degradation of the crystal structure. In contrast, $\alpha$-RuCl$_3$ exhibits a discontinuity and hysteresis in the in-plane lattice parameter at the structural transition, indicating that the stacking rearrangement is more strongly coupled to the intralayer bonding geometry. Similar to strain accommodation effects observed in martensitic transformations \cite{khachaturyan_theory_2013}, the abrupt in-plane lattice discontinuity may generate transient elastic strain during the first-order transition resulting in irreversible lattice inhomogeneity and reduced in-plane and out-of-plane structural coherence upon thermal cycling. This is consistent with the experimental observations of degradation of the crystalline quality after thermal cycles. 

The distinct structural reversibility of these two compounds is paralleled by different magnetic diffuse scattering behavior. In CrCl$_3$, magnetic diffuse scattering is observed well above $T_N$, forming rods along the out-of-plane direction that are characteristic of quasi-two-dimensional correlations with weak interlayer coupling \cite{osborn_diffuse_2025}. The gradual buildup of intensity near the antiferromagnetic propagation vector upon cooling suggests the development of strong short-range ferromagnetic correlations within the honeycomb layers, while the interlayer correlations remain only weakly correlated until three-dimensional long-range antiferromagnetic order emerges near $T_N$ \cite{liu_anisotropic_2020}. In contrast, no comparable quasi-static magnetic diffuse scattering is observed in $\alpha$-RuCl$_3$ within the sensitivity of our neutron diffraction measurements.

In order to understand the origin of the above contrasting behavior in both structural reversibility and magnetic diffuse features described above, both the lattice geometry and the electronic configurations of transition metal ions should be considered. Geometrically, the first order structural transition between the high-temperature $C2/m$ and low-temperature $R\bar{3}$ structures in both compounds involves lateral displacements of adjacent honeycomb layers. Differences in interlayer coupling and sliding energetics, lattice stiffness, and shear accommodation can influence how layer sliding  couples to the in-plane structure. However, the different in-plane lattice responses to the structure transition of these two compounds indicate that geometric consideration alone is unlikely to fully account for the contrast behavior. The fundamentally different electronic configurations of Cr$^{3+}$ $3d^3$ and Ru$^{3+}$ $4d^5$ could play an important role. 

CrCl$_3$ is a comparatively conventional $3d^3$ system with weak spin-orbit coupling and predominantly isotropic exchange interactions, whereas $\alpha$-RuCl$_3$ is a spin-orbit-entangled $4d^5$ Mott insulator in which the bond dependent exchange interactions are highly sensitive to the local bonding geometry. As a result, small variations in octahedral distortions can have a dramatic effect on the total electronic energy.  The stacking sequence change during the structure transition perturbs the bonding environment across the van der Waals gap, and the associated lattice distortions impacts the electronic energy more strongly in $\alpha$-RuCl$_3$ than in CrCl$_3$. Consequently, the stacking rearrangement couples more strongly to the in-plane lattice degrees of freedom in $\alpha$-RuCl$_3$, producing the observed discontinuous in-plane lattice response and crystalline degradation upon thermal cycling.

A similar strong coupling between the stacking rearrangement and an in-plane lattice response is observed in CrI$_3$ \cite{dolezal_formation_2024}. In that compound, thermal cycling leads to a history-dependent transformation upon cooling due to the formation of structural domains and mosaic blocks with varying defect densities. Compared to CrCl$_3$, CrI$_3$ has a stronger ligand spin-orbit coupling and covalency. This seems to support the idea that the spin-orbit assisted electronic interactions play an important role in governing the lattice response of these layered honeycomb halides. A comparative study of CrCl$_3$, CrBr$_3$, and CrI$_3$ would further clarify this. 

The different electronic configurations likely underlie the contrasting magnetic diffuse features observed in CrCl$_3$ and $\alpha$-RuCl$_3$. In CrCl$_3$, the weak spin-orbit coupling isotropic exchange interactions favor conventional quasi-two-dimensional magnetic correlations that gradually develop upon cooling toward T$_N$. These correlations lead to the well defined magnetic diffuse features observed in our neutron diffraction suggesting strong intralayer correlations coexisting with weak interlayer coupling. In $\alpha$-RuCl$_3$, strong spin-orbit coupling results in anisotropic and frustrated bond dependent magnetic interactions that tend to enhance magnetic fluctuations and suppress the formation of the short-range order above the zig-zag magnetic ordering \cite{laurell_dynamical_2020}. Spectroscopic studies in related Kitaev honeycomb iridates have further revealed broad magnetic excitation continua persisting to temperatures far above $T_N$ \cite{revelli_fingerprints_2020}, suggesting that the spin-orbit-entangled interactions underlying the high-temperature magnetic fluctuations may contribute to the pronounced in-plane lattice response and reduced structural reversibility observed in $\alpha$-RuCl$_3$.

Thermal cycling in $\alpha$-RuCl$_3$ does not appear to induce any measurable feature in heat capacity or magnetization even after many cycles \cite{zhang_stacking_2024}. On the other hand, the crystalline degradation induced by the in-plane lattice change across the structure transition might have important implications for the studies of thermal transport properties of $\alpha$-RuCl$_3$. A reduction in either in-plane or out-of-plane correlation lengths as well as increased mosaicity after thermal cyclings are expected to scatter the heat carriers, thereby influencing both the longitudinal thermal conductivity and the transverse thermal Hall response. These effects may further complicate the experimental realization and interpretation of the half-integer quantized thermal Hall effect that has shown strong sample dependence \cite{yamashita2020sample, kasahara2022quantized, zhang_sample-dependent_2023, zhang_stacking_2024}.

\section{Summary}

We have performed a comparative neutron scattering study of the layered honeycomb magnets CrCl$_3$ and $\alpha$-RuCl$_3$, revealing distinct structural reversibility and magnetic diffuse features. While both compounds undergo a stacking-driven structural transition, CrCl$_3$ exhibits a smooth in-plane lattice response and remains structurally robust under thermal cycling, whereas $\alpha$-RuCl$_3$ shows a discontinuous in-plane lattice change accompanied by progressive degradation of crystalline quality with a reduction of in-plane and out-of-plane coherence. These differences are further reflected in their magnetic behavior, with CrCl$_3$ exhibiting quasi-two-dimensional diffuse scattering above $T_N$, while $\alpha$-RuCl$_3$ shows no comparable quasi-static correlations within experimental sensitivity. 

These contrasting behaviors likely originate from the fundamentally different electronic character of the two compounds. Compared to the weak spin-orbit coupling and predominantly isotropic interactions in CrCl$_3$, the spin-orbit-entangled electronic state and bond-dependent interactions in $\alpha$-RuCl$_3$ enhance its sensitivity to local bonding geometry, which leads to a stronger coupling between layer sliding and in-plane lattice degrees of freedom.

\begin{acknowledgments}
IY was supported by an appointment to the Oak Ridge National Laboratory (ORNL)
Next Generation Pathway to Computing Program (NGP) Program,
sponsored by the U.S. Department of Energy and administered by the Oak Ridge Institute for Science and Education. MAM and JQY were supported by the U.S. Department of Energy, Office of Science, Basic Energy Sciences, Materials Sciences and Engineering Division. This research used resources at the Spallation Neutron Source, a DOE Office of Science User Facility operated by ORNL. The beam time was allocated to CORELLI on proposal numbers IPTS-33613.1 and IPTS-31238.4. All drawings of crystal and magnetic structures were obtained using VESTA software \cite{momma_vesta_2011}. This manuscript has been authored by UT-Battelle, LLC, under Contract No. DE-AC0500OR22725 with the U.S. Department of Energy.
\end{acknowledgments}

\bibliography{honeycomb_reduced,supplemental_reduced}





\end{document}